\documentclass[review,sort&compress,numbers,longnamesfirst]{elsarticle}
\usepackage{graphicx,floatrow,subfigure}
\usepackage{float,fancyhdr,bm,caption}
\usepackage{amssymb,amsmath,amsthm,wasysym,indentfirst}
\usepackage{graphicx,accents,statex}
\usepackage{latexsym,bm}
\usepackage{amsthm,holtpolt,stfloats,CJK}
\usepackage[all]{xy}
\usepackage{indentfirst,enumerate,caption,mdframed}
\usepackage{amsmath}
\usepackage{longtable}
\usepackage{colortab}
\usepackage{enumerate}
\usepackage{lineno,hyperref}
\usepackage{geometry}
\journal{ }

\newcommand{\supercite}[1]{\textsuperscript{\cite{#1}}}
\theoremstyle{definition}
\newtheorem{definition}{\hskip 1.3em Definition}
\newtheorem{example}{\hskip 1.3em Example}
\newtheorem{theorem}{\hskip 1.3em Theorem}
\newtheorem{lemma}{\hskip 1.3em Lemma}

\usepackage{paralist}

\allowdisplaybreaks

\usepackage{array}
\newcommand{\PreserveBackslash}[1]{\let\temp=\\#1\let\\=\temp}
\newcolumntype{C}[1]{>{\PreserveBackslash\centering}p{#1}}
\newcolumntype{R}[1]{>{\PreserveBackslash\raggedleft}p{#1}}
\newcolumntype{L}[1]{>{\PreserveBackslash\raggedright}p{#1}}

\makeatletter
\def\wideubar{\underaccent{{\cc@style\underline{\mskip10mu}}}}
\def\Wideubar{\underaccent{{\cc@style\underline{\mskip8mu}}}}
\makeatother

\makeatletter
\def\widebar{\accentset{{\cc@style\underline{\mskip10mu}}}}
\def\Widebar{\accentset{{\cc@style\underline{\mskip8mu}}}}
\makeatother

\makeatletter
\renewcommand*\env@matrix[1][\arraystretch]{%
  \edef\arraystretch{#1}%
  \hskip -\arraycolsep
  \let\@ifnextchar\new@ifnextchar
  \array{*\c@MaxMatrixCols c}}
\makeatother

\begin{document}
\begin{CJK}{GBK}{song}
\begin{frontmatter}
\title{Reconstruct the Logical Network from the Transition Matrix\tnoteref{mytitlenote}}
\tnotetext[mytitlenote]{This work was supported by National Natura Science Foundation of China under Grants 60774007 and 61305101.}

\author{Cailu Wang}

\author{Yuegang Tao\corref{mycorrespondingauthor}}
\cortext[mycorrespondingauthor]{Corresponding author}
\ead{yuegangtao@hebut.edu.cn}

\address{School of Control Science and Engineering, Hebei University of Technology, Tianjin, 300130, P. R. China}

\begin{abstract}
Reconstructing the logical network from the
transition matrix is benefit for learning the logical meaning of the algebraic result from the
algebraic representation of a BN. And so far there has no method to convert the matrix expression back to the logic expression for a BN with an
arbitrary topology structure.
Based on the canonical form and Karnaugh map, we propose a
method for
reconstructing the logical network from the transition matrix of a Boolean network in this paper.
\end{abstract}


\end{frontmatter}

\section{Introduction}
The Boolean network (BN) is a logical system pioneered by  Kauffman in \cite{Kauffman} for describing genetic regulatory networks. BNs provide useful modeling tools for dynamical systems whose state-variables can attain two
possible values, and have
attracted a considerable attention in computer networks \cite{BooleanFunctions}, gene networks \cite{NetworkBiology,LAC}, neural networks \cite{Attractor,RetrievingBoolean},
as well as social networks \cite{SocialNetwork}.

A matrix
product, called the semi-tensor product of matrices, is developed in \cite{LinearRepresentation} to deal with the BNs.
Using this
method, a BN can be modeled by a standard discrete-time linear system.
Converting a logical dynamic system
into a standard discrete-time linear system is an indispensable key step in analysing dynamics of BNs through the algebraic method.
Conversely, reconstructing the logical network from the transition matrix is advantageous for learning the logical meaning of the algebraic result from the algebraic representation of a BN.
Under the premise that the in-degrees of nodes are smaller than the total number of nodes, some formulas to retrieve the logical dynamic equations from the transition matrix is provided in \cite{Controllabilityobservability}.
However,
there do exist an amount of BNs with nodes whose neighborhood contains all nodes.
It is easy to give some examples.

\begin{center}
\vbox{
\ \includegraphics[width=1.6in]{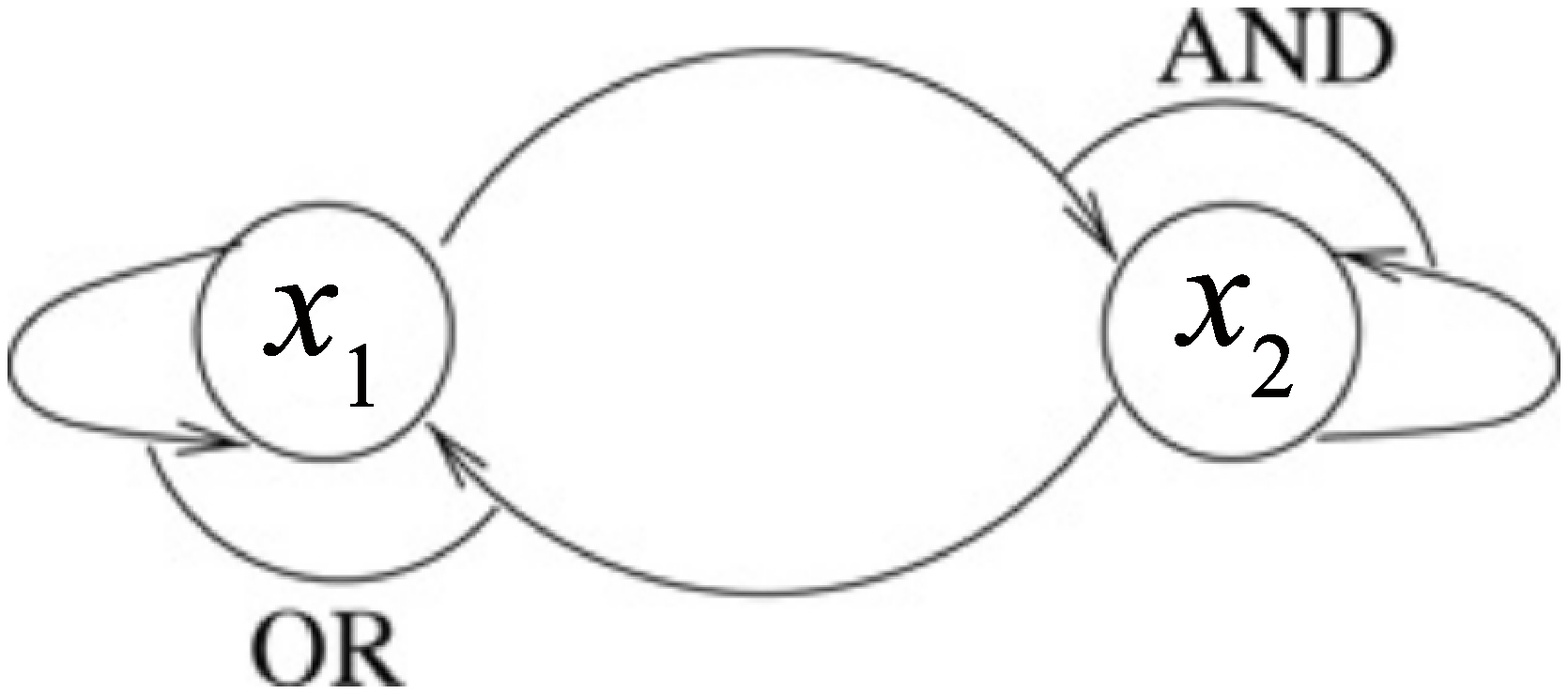}\ \ \ \ \ \ \ \ \ \ \ \ \ \includegraphics[width=1.6in]{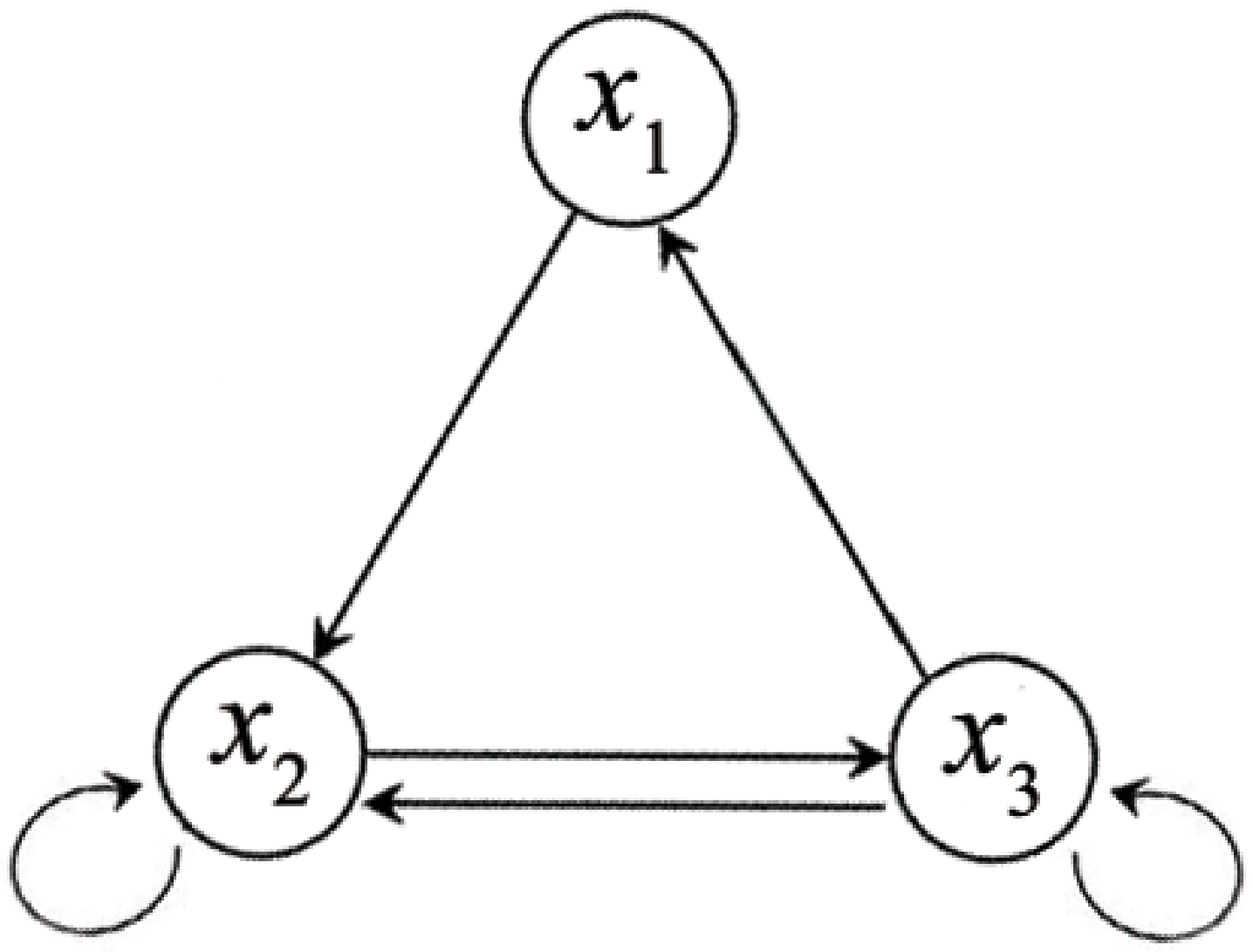}\vskip0mm {\small\ \ \ \ \ \ \
Fig.\,1\supercite{Controllabilityobservability}. Logical network\ \ \ \ \ \ \ \  \ \ \ \ \ \ \ \ \ \
Fig.\,2. Logical network\ \ }
}
\end{center}

\begin{example}
Let us see the BN given in \cite[Fig.1]{Controllabilityobservability}.
Figure 1 is a BN with 2 nodes,
in which the in-degrees of nodes $x_{1}$ and $x_{2}$ are both equal to 2.
Hence, the logic expression of both two nodes $x_{1}$ and $x_{2}$ can not be retrieved from the matrix expression by using the conversion formulas provided in \cite{Controllabilityobservability}.
\end{example}

\begin{example}
Figure 2 is a BN with 3 nodes,
in which the in-degrees of node $x_{2}$ is equal to 2. Hence, the logic expression of node $x_{2}$ can not be retrieved from the matrix expression by using the conversion formulas provided in \cite{Controllabilityobservability}.
\end{example}

In a word, the method proposed in \cite{Controllabilityobservability} can not reconstruct the logical networks from the transition matrices for such a type of BNs
(see e.g. Example \ref{ex2} in this paper).
A natural question may be proposed: How to convert the matrix expression back to the logic expression for a BN with an arbitrary topology structure?
This paper will give an answer to it.


The theory and applications of the canonical form in
mathematics, computer sciences and logic have been investigated by many researchers. For instance, Gunawardena \cite{Minmaxfunctions} presented the {canonical form} of a max-min system introduced in \cite{dynamicmaxmin}.
Based on such a {canonical form}, some interesting results with profound significance have been obtained in max-min systems,
such as the duality theorem \cite{dualitytheorem} and constructive
fixed point theorem \cite{fixedpoint}, and
some control problems are also considered (see e.g. \cite{DEDS,StateFeedback,112}).
It is known that any Boolean function can be put into {a} minterm canonical form \cite{BooleanFunctions}.
Based on such a canonical form,
we will give a direct conversion from the logic expression to the
matrix expression of a BN in this paper.

The Karnaugh map (K-map) is a useful tool to simplify the logic expressions \cite{LogicDesign}. In this paper, we will introduce two types of K-maps --
 the K-map
of a BN and the K-map of a logical matrix -- to convert the matrix expression back to the logic expression.
It is proven that
the K-map of
a logical matrix is exactly the K-map of the BN determined by this logical matrix.
Then, we can convert
the
matrix expression back to a canonical form of the logic
expression through plotting the K-map of the transition matrix.

The outline of the rest of {this paper} is organized as follows.
Section 2 introduces some basic concepts about BNs, and points out an unsolved problem existing in the linear representation theory of BNs.
Such an unsolved problem is solved in Section 3. Some future works are drawn in
Section 4.

\section{Preliminary and problem description}
A \emph{Boolean function} is a function of the form $$f:\Delta^{n}\rightarrow \Delta,$$ where $\Delta=\{\text{True},\text{False}\}$ is the logical domain and $n\in\mathbb{N}$ (the set of natural numbers) is the arity of the function. Any $n$-variable Boolean function can be expressed as a propositional formula in variables $x_{1},x_{2},\cdots,x_{n}$.
For simplicity of
presentation,
the logical disjunction $A\vee B$, the logical conjunction $A\wedge B$ and the logical negation $\neg A$ are represented by $A+B$, $AB$ and $\widebar{A}$, respectively.

\begin{table}[H]
  \centering
  \begin{tabular}{|c|ccc|c|}
     \hline
     Row No. & $x_{1}$ &$x_{2}$ &$x_{3}$ & Minterms \\ \hline
     0 & 0&0&0 & $\widebar{x}_{1}\widebar{x}_{2}\widebar{x}_{3}=m_{0}$ \\
     1 & 0&0&1 & $\widebar{x}_{1}\widebar{x}_{2}{x_{3}}=m_{1}$ \\
     2 & 0&1&0 & $\widebar{x}_{1}{x_{2}}\widebar{x}_{3}=m_{2}$ \\
     3 & 0&1&1 & $\widebar{x}_{1}{x_{2}}{x_{3}}=m_{3}$ \\
     4 & 1&0&0 & ${x_{1}}\widebar{x}_{2}\widebar{x}_{3}=m_{4}$ \\
     5 & 1&0&1 & ${x_{1}}\widebar{x}_{2}{x_{3}}=m_{5}$ \\
     6 & 1&1&0 & ${x_{1}}{x_{2}}\widebar{x}_{3}=m_{6}$ \\
     7 & 1&1&1 & ${x_{1}}{x_{2}}{x_{3}}=m_{7}$ \\
     \hline
   \end{tabular}
  \caption{Minterms for a 3-variable Boolean function}
\end{table}

A \emph{truth table} specifies the values of a
Boolean function for every possible combination of values of the variables in the function.
A \emph{minterm}
of $n$ variables is a product of $n$ literals in which each variable appears exactly once
in either true or complemented form, but not both.
In general, the minterm which corresponds to the row number $i$ of
the truth table is designated $m_{i}$ ($i$ is usually written in decimal).
Table 1 lists all of the minterms of a 3-variable Boolean function. Any Boolean function $f$ can uniquely be written as a sum of minterms, i.e.,
\begin{equation*}
 f(x_{1},x_{2},\cdots,x_{n})=m_{i_{1}}+m_{i_{2}}+\cdots+m_{i_{k}}:=\sum m(i_{1},i_{2},\cdots,i_{k}),
\end{equation*}
which
is called the \emph{minterm {canonical form}} of $f$. Given the minterm {canonical form} of a Boolean function, it can be plotted on a \emph{Karnaugh map} (K-map) by placing 1 in the squares which correspond to minterms of the function and 0 in the
remaining squares.
Figure 1 shows the location of
minterms on a 3-variable K-map.

\begin{center}
\vbox{
\includegraphics[width=1.8in]{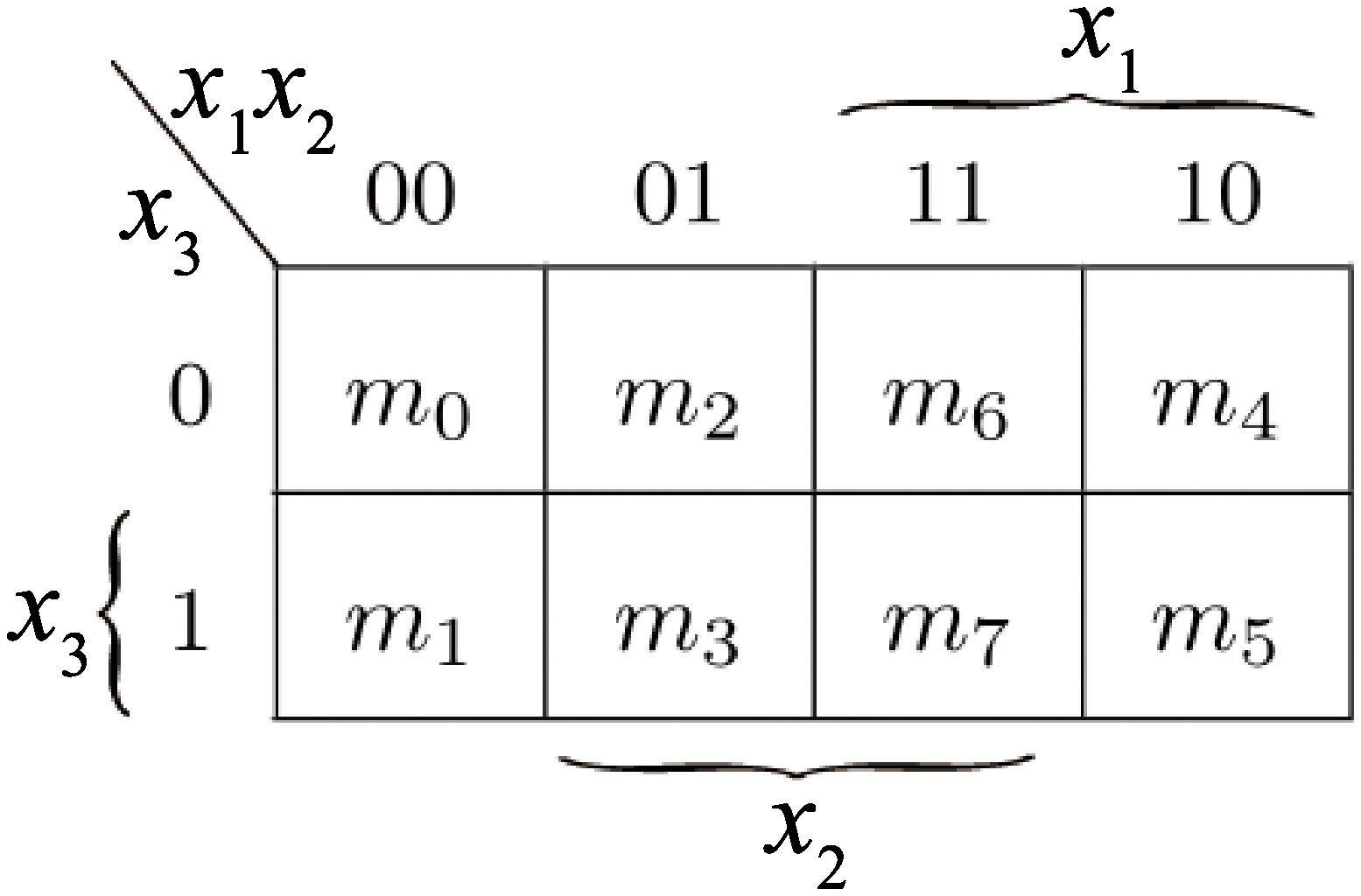}
\vskip-2mm{\small Fig.\,3. Location of minterms on a 3-variable K-map}
}
\end{center}

A \emph{Boolean network} (BN) with a set of nodes $x_{1}(t)$, $x_{2}(t)$, $\cdots$, $x_{n}(t)$
can be described as
\begin{equation}\label{e7}
\left\{
\begin{array}{lll}
x_{1}(t+1)&=&f_{1}\big(x_{1}(t),x_{2}(t),\cdots,x_{n}(t)\big),\\
x_{2}(t+1)&=&f_{2}\big(x_{1}(t),x_{2}(t),\cdots,x_{n}(t)\big),\\
\multicolumn{3}{c}{\dotfill}\\
x_{n}(t+1)&=&f_{n}\big(x_{1}(t),x_{2}(t),\cdots,x_{n}(t)\big),
\end{array}\right.
\end{equation}
where $t\in\mathbb{N}$ (the set of natural numbers) and $f_{i}$ $(i=1,2,\cdots,n)$ are Boolean functions.

\begin{definition}\label{d1}\supercite{LinearRepresentation}
Let $A\in \mathbb{R}^{m\times n}$ and $B\in \mathbb{R}^{p\times q}$. The \emph{semi-tensor product} of  $A$ and $B$ is defined as
$$A\ltimes B=(A\otimes I_{\alpha/n})\times(B\otimes I_{\alpha/p}),$$ where $\alpha$ is the  least common multiple of $n$ and $p$, $I$ is the identity matrix, $\otimes$ is the Kronecker product and  $\times$ is the conventional product of matrices.
\end{definition}

Using the semi-tensor product of matrices, a Boolean function can be converted
into an algebraic form.
Let $\delta_{n}^{i}$
denote the $i$-th column of the identity matrix $I_{n}$.
For the sake of brevity,  a matrix of the form $L=[\delta_{n}^{i_{1}}\ \delta_{n}^{i_{2}}\ \cdots\ \delta_{n}^{i_{r}}]$ is briefly denoted as $L=\delta_{n}[i_{1}, i_{2}, \cdots, i_{r}]$.
Represent
the logical values ``True" and ``False" by
$$\delta_{2}^{1}=\begin{bmatrix}
1\\
0
\end{bmatrix}\ \text{and}\
\delta_{2}^{2}=\begin{bmatrix}
0\\
1
\end{bmatrix},
$$respectively.
Then any $n$-variable Boolean function
can be equivalently represented as
\begin{equation*}
f: \{\delta_{2}^{1}, \delta_{2}^{2}\}^{n}\rightarrow
\{\delta_{2}^{1}, \delta_{2}^{2}\}.
\end{equation*}

\begin{lemma}\supercite{LinearRepresentation}
Let $x(t)=x_{1}(t)\ltimes x_{2}(t)\ltimes\cdots \ltimes x_{n}(t)$. Then
there exists a unique matrix $L\in\{0,1\}^{2^{n}\times2^{n}}$ such that BN (\ref{e7}) can be converted
to a standard discrete-time dynamic system as
\begin{equation}\label{e8}
x(t+1)=L\ltimes x(t),\ t=0,1,\cdots,
\end{equation}
where $L$ is called the \emph{transition matrix} of BN (\ref{e7}).
\end{lemma}

A \emph{logical matrix} is a binary matrix of the form \begin{equation}\label{e3}
\delta_{2^{n}}[i_{1},i_{2},\cdots,i_{2^{n}}],
\end{equation} where $i_{r}\in\{0,1,\cdots,2^{n}-1\}$ and $r\in\{1,2,\cdots,2^{n}\}$.
There exists a one-to-one correspondence between
logical matrices and BNs.

The following method is provided in  \cite{Controllabilityobservability} to retrieve
the logical network from the transition matrix of a BN, in the condition that the in-degree of a node is smaller
than the total number of nodes in the BN. Defined a set of $2 \times 2^{n}$ matrices $S_{i}^{n}$ as follows:
\begin{align*}
S_{1}^{n}&=\delta_{2}[\underbrace{1,\cdots,1}_{2^{n-1}},\underbrace{2,\cdots,2}_{2^{n-1}}],\\
S_{2}^{n}&=\delta_{2}[\underbrace{1,\cdots,1}_{2^{n-2}},\underbrace{2,\cdots,2}_{2^{n-2}},\underbrace{1,\cdots,1}_{2^{n-2}},\underbrace{2,\cdots,2}_{2^{n-2}}],\\
\vdots&\ \\
S_{n}^{n}&=\delta_{2}[\underbrace{1,2,1,2,\cdots,1,2}_{2^{n}}].
\end{align*}

\begin{lemma}\supercite{Controllabilityobservability}\label{l3}
For a BN with its matrix expression (\ref{e8}),
the structure matrix of $f_{i}$ can be retrieved as $$M_{i}=S_{i}^{n}\ltimes L,\ i=1,2,\cdots,n.$$
If $M_{i}$ satisfies
\begin{equation}\label{e11}
  M_{i}\ltimes W_{[2,2^{j-1}]}\ltimes(M_{\neg}-I_{2})=0,\ 1\leq j\leq n,
\end{equation}
where $M_{\neg}=\delta_{2}[2,1]$ is the structure matrix of negation $\neg$ and $W_{[*,*]}$ is the swap matrix,
then the equation of $f_{i}$ can be replaced by
\begin{equation}\label{e2}
  x_{i}(t+1)=M_{i}'\ltimes x_{1}(t)\ltimes\cdots\ltimes x_{j-1}(t)\ltimes x_{j+1}(t)\ltimes\cdots \ltimes x_{n}(t),
\end{equation}
where $M_{i}'=M_{i}\ltimes W_{[2,2^{j-1}]}\ltimes \delta_{2}^{1}$.
\end{lemma}

The following is a numeral example of a BN whose logic network can not be reconstructed from its matrix expression by using the method above.

\begin{example}\label{ex2}
Reconstruct the logic network of the BN
\begin{equation}\label{e13}
  x(t+1)=L\ltimes x(t),\ t=0,1,\cdots,
\end{equation}
where $L=\delta_{4}[1,2,2,4]$. By a direct calculation,
\begin{align*}
  M_{1}&=S_{1}^{2}\ltimes L=\delta_{2}[1,1,1,2],\\
   M_{2}&=S_{2}^{2}\ltimes L=\delta_{2}[1,2,2,2].
\end{align*}
For $i=1$, it can be verified that
\begin{align*}
  M_{1}\ltimes W_{[2,1]}\ltimes(M_{\neg}-I_{2})&=\begin{bmatrix}0&-1&0&1\\0&1&0&-1\end{bmatrix}\neq 0,\\
  M_{1}\ltimes W_{[2,2]}\ltimes(M_{\neg}-I_{2})&=\begin{bmatrix}0&-1&0&1\\0&1&0&-1\end{bmatrix}\neq 0.
\end{align*}
Then equation (\ref{e11}) does not hold for any $j\in\{1,2\}$, and hence the logic expression of $f_{1}$ can not be reconstructed by using (\ref{e2}).
For $i=2$, it can be verified that
\begin{align*}
  M_{2}\ltimes W_{[2,1]}\ltimes(M_{\neg}-I_{2})&=\begin{bmatrix}-1&0&1&0\\1&0&-1&0\end{bmatrix}\neq 0,\\
  M_{2}\ltimes W_{[2,2]}\ltimes(M_{\neg}-I_{2})&=\begin{bmatrix}-1&0&1&0\\1&0&-1&0\end{bmatrix}\neq 0.
\end{align*}
Then equation (\ref{e11}) does not hold for any $j\in\{1,2\}$, and hence the logic expression of $f_{2}$ also can not be reconstructed by using (\ref{e2}).
\end{example}

Now that the logical network of BN (\ref{e13}) can not be reconstructed by using Lemma \ref{l3},
a natural question may be proposed: How to reconstruct the logical network of a BN with an arbitrary topological structure?

\section{Problem solving}
In
this
section, we will provide a solution method for the unsolved problem of
reconstructing the logical network from the transition matrix of a BN through
introducing the following two types of K-maps.

\begin{definition}\label{d2}
The \emph{K-map of BN} (\ref{e7}) is generated by composing the K-maps of all Boolean functions $f_{r}$ into a new map, in which the cell value in the square $k$ is the decimal representation of the binary number whose $r$-th bit is the cell value in the square $k$ of the K-map of $f_{r}$, where $1\leq r\leq n$.
\end{definition}

\begin{definition}
The \emph{K-map of logical matrix} (\ref{e3}) is defined by a K-map with
\begin{equation}\label{e12}
d_{k}=2^{n}-i_{2^{n}-k},\ k=0,1,\cdots, 2^{n}-1,
\end{equation} as the cell value in the square $k$.
\end{definition}

Next, let us review the relationship between the above two K-maps.

\begin{lemma}\label{l2}
Let $i_{r}\in \{1,2\}\ (r=1,2,\cdots,n)$. Then
\begin{equation}\label{e5}
\delta_{2}^{i_{1}}\ltimes \delta_{2}^{i_{2}}\ltimes\cdots\ltimes \delta_{2}^{i_{n}}=\delta_{2^{n}}^{2^{n}-a(n)},
\end{equation}
where
\begin{equation}\label{e9}
  a(n)=\sum_{r=1}^{n}(i_{r} \text{\ Mod\ } 2)\times2^{n-r}.
\end{equation}
\end{lemma}
\emph{Proof:}
Let us use the induction on $n$ to prove this lemma.
It is easy to calculate that
$i_{1}=2-(i_{1}\text{\ Mod\ } 2)$ for $i_{1}=1,2$.
Hence $\delta_{2}^{i_{1}}=\delta_{2}^{2-(i_{1}\text{\ Mod\ } 2)}$, i.e., (\ref{e5}) holds for $n=1$. Assume that (\ref{e5}) holds for $n=k$.
Then, for $n=k+1$,
$$\delta_{2}^{i_{1}}\ltimes \delta_{2}^{i_{2}}\ltimes\cdots\ltimes \delta_{2}^{i_{k}}\ltimes \delta_{2}^{i_{k+1}}=\delta_{2^{k}}^{2^{k}-a(k)}\ltimes \delta_{2}^{i_{k+1}}$$
by the inductive hypothesis.
Denote $l=2^{k}-a(k)$. Consider which component of $\delta_{2^{k}}^{l}\ltimes \delta_{2}^{i_{k+1}}$ is 1.
Since the $l$-th  component of $\delta_{2^{k}}^{l}$ is 1, the first $2(l-1)$ components of $\delta_{2^{k}}^{l}\ltimes \delta_{2}^{i_{k+1}}$ are 0s
according to Definition \ref{d1}.
If $i_{k+1}=1$, then $(i_{k+1} \text{\ Mod\ } 2)=1$ and the $2l-1\big(=2(l-1)+1\big)$-th  component of $\delta_{2^{k}}^{l}\ltimes \delta_{2}^{i_{k+1}}$ is 1. Since
\begin{align*}
2l-1&=2^{k+1}-2a(k)-1\\
&=2^{k+1}-2\sum_{r=1}^{k}(i_{r} \text{\ Mod\ } 2)\times2^{k-r}-(i_{k+1} \text{\ Mod\ } 2)\\
&=2^{k+1}-\sum_{r=1}^{k+1}(i_{r} \text{\ Mod\ } 2)\times2^{k+1-r}\\
&=2^{k+1}-a(k+1),
\end{align*}
we have
$$\delta_{2^{k}}^{l}\ltimes \delta_{2}^{i_{k+1}}=\delta_{2^{k+1}}^{2^{k+1}-a(k+1)}.$$
Similarly, the above equation also holds for $i_{k+1}=2$.
Hence (\ref{e5}) holds for $n=k+1$, and so it holds for all positive integers $n$.
\qed\vspace{1.5mm}

\begin{lemma}\label{t3}
Let $c_{k}\ (0\leq k \leq 2^{n}-1)$ be the number in the square $k$ of the K-map of BN (\ref{e7}). Then the transition matrix of BN (\ref{e7}) is given by
\begin{equation}\label{e1}
L=\delta_{2^{n}}[2^{n}-c_{2^{n}-1},\cdots,2^{n}-c_{1},2^{n}-c_{0}].
\end{equation}\end{lemma}
\emph{Proof:}
For BN (\ref{e7}), let
$x_{r}(t)=\delta_{2}^{i_{r}}$, where $i_{r}\in \{1,2\}$ and $r=1,2,\cdots,n$.
It follows from Lemma \ref{l2} that
 $$x(t)=\ltimes_{r=1}^{n}x_{r}(t)=\ltimes_{r=1}^{n}\delta_{2}^{i_{r}}=\delta_{2^{n}}^{2^{n}-k},$$
where $k$ is given in (\ref{e9}). Let
$x_{r}(t+1)=\delta_{2}^{j_{r}}\ (j_{r}\in \{1,2\})$. It follows from Lemma \ref{l2} that
$$x(t+1)=\ltimes_{r=1}^{n}x_{r}(t+1)=\ltimes_{r=1}^{n}\delta_{2}^{j_{r}}=\delta_{2^{n}}^{2^{n}-c_{k}},$$
where $$c_{k}=\sum_{r=1}^{n}(j_{r} \text{\ Mod\ } 2)\times2^{n-r}$$ is just the number in the square $k$ of the K-map of BN (\ref{e7}).
Since $L\ltimes x(t)=L\ltimes \delta_{2^{n}}^{2^{n}-k}$ represents the $(k+1)$-th column of matrix $L$ from the right, one obtains $$L\ltimes x(t)=\delta_{2^{n}}^{2^{n}-c_{k}}=x(t+1).$$
Hence, $L$ is the transition matrix of BN (\ref{e7}).
\qed

\begin{theorem}
The K-map of
a logical matrix is  the K-map of the BN determined by this logical matrix.
\end{theorem}
\emph{Proof:}
Let $L=\delta_{2^{n}}[i_{1}\ i_{2}\ \cdots\ i_{2^{n}}]$ be
a logical matrix, and
$d_{k}$ be the cell value in the square $k$ of the K-map of $L$.
Let $c_{k}$ be the cell value in the square $k$ of the K-map of BN (\ref{e8}), whose transition matrix is $L$.
By (\ref{e1}), $$i_{r}=2^{n}-c_{2^{n}-r},\ r=1,2,\cdots,2^{n}.$$
Let $k=2^{n}-r$. Then $$c_{k}=2^{n}-i_{2^{n}-k}=d_{k},\ k=0,1,\cdots,2^{n}-i.$$
This implies that the K-map of $L$ is the K-map of BN (\ref{e8}).
\qed\vspace{1.5mm}

Owing to the theorem above,
the following operation
procedure is developed to convert the matrix expression back to the logic expression of BN (\ref{e8}).
\begin{enumerate}[Step 1.]
  \item Plot the K-map of $L$ by formula (\ref{e12}).
  \item Plot the K-map of ${f_{r}}$ for each $r\in\{1,2,\cdots,n\}$, in which the cell value in the square $k$ is the $r$-th bit of the binary representation
    of the decimal number $d_{k}$.
  \item Construct the minterm canonical form of  each ${f_{r}}$ from its K-map.\vspace{1.5mm}
\end{enumerate}

By using the method above, the unsolved problem in Example \ref{ex2} can be solved.

\begin{example}\label{ex1}
Reconstruct the logic network of the BN (\ref{e13}).
\begin{enumerate}[Step 1]
  \item Plot the K-map of $L$. By (\ref{e12}),
  \begin{align*}
d_{0}=4-i_{4}=4-4=0,\ &d_{1}=4-i_{3}=4-2=2,\\
d_{2}=4-i_{2}=4-2=2,\ &d_{3}=4-i_{1}=4-1=3.
\end{align*}
Then the K-map of BN (\ref{e13}) is given as follows:
\begin{center}
\vbox{
\renewcommand\arraystretch{1.3}
    \begin{tabular}{R{0.4cm}|C{0.4cm}|C{0.4cm}|}
    \multicolumn{1}{r}{}
    &\multicolumn{1}{c}{{\scriptsize 0}}&\multicolumn{1}{c}{{\scriptsize 1}}\\ \cline{2-3}
    {\scriptsize 0} &0&2\\ \cline{2-3}
    {\scriptsize 1} &2&3\\ \cline{2-3}
     \end{tabular}\ \ \ \ \ \ \ \ \ \ \ \ \ \begin{tabular}{R{0.4cm}|C{0.4cm}|C{0.4cm}|}
    \multicolumn{1}{r}{}
    &\multicolumn{1}{c}{{\scriptsize 0}}&\multicolumn{1}{c}{{\scriptsize 1}}\\ \cline{2-3}
    {\scriptsize 0} &00&10\\ \cline{2-3}
    {\scriptsize 1} &10&11\\ \cline{2-3}
     \end{tabular}\vskip2mm {\small\ \ \ \ \ \ \
(a) Decimal notation\ \ \ \ \ \  \ \ \ \ \ \ \
(b) Binary notation}
\vskip2mm{\small\ \ \ \ \ Fig.\,4. K-map of BN (\ref{e13})}\vspace{1.3mm}
}
\end{center}

  \item Plot the K-maps of $f_{1}$ and $f_{2}$.
\begin{center}
\vbox{
\renewcommand\arraystretch{1.3}
    \begin{tabular}{R{0.4cm}|C{0.4cm}|C{0.4cm}|}
    \multicolumn{1}{r}{}
    &\multicolumn{1}{c}{{\scriptsize 0}}&\multicolumn{1}{c}{{\scriptsize 1}}\\ \cline{2-3}
    {\scriptsize 0} &0&1\\ \cline{2-3}
    {\scriptsize 1} &1&1\\ \cline{2-3}
     \end{tabular}\ \ \ \ \ \ \ \ \ \ \ \begin{tabular}{R{0.4cm}|C{0.4cm}|C{0.4cm}|}
    \multicolumn{1}{r}{}
    &\multicolumn{1}{c}{{\scriptsize 0}}&\multicolumn{1}{c}{{\scriptsize 1}}\\ \cline{2-3}
    {\scriptsize 0} &0&0\\ \cline{2-3}
    {\scriptsize 1} &0&1\\ \cline{2-3}
     \end{tabular}\vskip2mm {\small\ \ \ \ \ \ \
(a) K-map of $f_{1}$\ \ \ \ \ \  \ \ \ \ \ \ \
(b) K-map of $f_{2}$}
\vskip2mm{\small\ \ \ \ \ Fig.\,5. K-map of Boolean functions}\vspace{1.3mm}
}
\end{center}

  \item The minterm canonical form of the logic expression is
 $$\left\{
\begin{array}{rcl}
x_{1}(t+1)&=&\sum m(1,2,3),\\
x_{2}(t+1)&=&m_{3}.
\end{array}\right.$$
Furthermore, the  minimization form is
 $$\left\{
\begin{array}{rcl}
x_{1}(t+1)&=&x_{1}(t)+x_{2}(t),\\
x_{2}(t+1)&=&x_{1}(t)x_{2}(t),
\end{array}\right.$$ and the network is presented in {Figure} 1.
\end{enumerate}
\end{example}

Let us see another example.

\begin{example}\label{ex1}
Reconstruct the logic network of the BN
\begin{equation}\label{e130}
  x(t+1)=L\ltimes x(t),\ t=0,1,\cdots,
\end{equation}
where $L=\delta_{8}[5,2,6,2,5,2,6,4]$.\vspace{1.5mm}
\begin{enumerate}[Step 1]
  \item Plot the K-map of $L$. By (\ref{e12}),
  \begin{align*}
d_{0}=8-i_{8}=8-4=4,\ &d_{1}=8-i_{7}=8-6=2,\\
d_{2}=8-i_{6}=8-2=6,\ &d_{3}=8-i_{5}=8-5=3,\\
d_{4}=8-i_{4}=8-2=6,\ &d_{5}=8-i_{3}=8-6=2,\\
d_{6}=8-i_{2}=8-2=6,\ &d_{7}=8-i_{1}=8-5=3.
\end{align*}
Then the K-map of BN (\ref{e130}) is given as follows:
\begin{center}
\vbox{
\renewcommand\arraystretch{1.3}
    \begin{tabular}{R{0.4cm}|C{0.4cm}|C{0.4cm}|C{0.4cm}|C{0.4cm}|}
    \multicolumn{1}{r}{}
    &\multicolumn{1}{c}{{\scriptsize 00}}&\multicolumn{1}{c}{{\scriptsize 01}}&\multicolumn{1}{c}{{\scriptsize 11}}&\multicolumn{1}{c}{{\scriptsize 10}}\\ \cline{2-5}
    {\scriptsize 0} &4&6&6&6\\ \cline{2-5}
    {\scriptsize 1} &2&3&3&2\\ \cline{2-5}
     \end{tabular}\ \ \ \ \ \ \ \ \ \ \ \ \begin{tabular}{R{0.4cm}|C{0.4cm}|C{0.4cm}|C{0.4cm}|C{0.4cm}|}
    \multicolumn{1}{r}{}
    &\multicolumn{1}{c}{{\scriptsize 00}}&\multicolumn{1}{c}{{\scriptsize 01}}&\multicolumn{1}{c}{{\scriptsize 11}}&\multicolumn{1}{c}{{\scriptsize 10}}\\ \cline{2-5}
    {\scriptsize 0} &100&110&110&110\\ \cline{2-5}
    {\scriptsize 1} &010&011&011&010\\ \cline{2-5}
     \end{tabular}\vskip2mm {\small\ \ \ \ \ \ \ \
(a) Decimal notation\ \ \ \ \ \  \ \ \ \ \ \ \ \ \ \ \ \ \ \ \ \ \ \ \ \
(b) Binary notation}
\vskip2mm{\small\ \ \ \ \ Fig.\,6. K-map of BN (\ref{e130})}\vspace{1.3mm}
}
\end{center}

  \item Plot the K-map of ${f_{r}}$ for each $r\in\{1,2,3\}$.
  \begin{center}{\centering
\vbox{\renewcommand\arraystretch{1.3}
\begin{tabular}{R{0.4cm}|C{0.4cm}|C{0.4cm}|C{0.4cm}|C{0.4cm}|}
    \multicolumn{1}{r}{}
    &\multicolumn{1}{c}{{\scriptsize 00}}&\multicolumn{1}{c}{{\scriptsize 01}}&\multicolumn{1}{c}{{\scriptsize 11}}&\multicolumn{1}{c}{{\scriptsize 10}}\\ \cline{2-5}
    {\scriptsize 0} &1&1&1&1\\ \cline{2-5}
    {\scriptsize 1} &0&0&0&0\\ \cline{2-5}
    \end{tabular}
    \vskip2mm {\small\ \ \ \ \ \ \
(a) K-map of $f_{1}$}
}}\end{center}
\begin{center}{\centering
\vbox{\renewcommand\arraystretch{1.3}
\begin{tabular}{R{0.4cm}|C{0.4cm}|C{0.4cm}|C{0.4cm}|C{0.4cm}|}
    \multicolumn{1}{r}{}
    &\multicolumn{1}{c}{{\scriptsize 00}}&\multicolumn{1}{c}{{\scriptsize 01}}&\multicolumn{1}{c}{{\scriptsize 11}}&\multicolumn{1}{c}{{\scriptsize 10}}\\ \cline{2-5}
    {\scriptsize 0} &0&1&1&1\\ \cline{2-5}
    {\scriptsize 1} &1&1&1&1\\ \cline{2-5}
    \end{tabular}
    \vskip2mm {\small\ \ \ \ \ \ \
(b) K-map of $f_{2}$}
}}\end{center}
\begin{center}{\centering
\vbox{\renewcommand\arraystretch{1.3}
\begin{tabular}{R{0.4cm}|C{0.4cm}|C{0.4cm}|C{0.4cm}|C{0.4cm}|}
    \multicolumn{1}{r}{}
    &\multicolumn{1}{c}{{\scriptsize 00}}&\multicolumn{1}{c}{{\scriptsize 01}}&\multicolumn{1}{c}{{\scriptsize 11}}&\multicolumn{1}{c}{{\scriptsize 10}}\\ \cline{2-5}
    {\scriptsize 0} &0&0&0&0\\ \cline{2-5}
    {\scriptsize 1} &0&1&1&0\\ \cline{2-5}
     \end{tabular}
    \vskip2mm {\small\ \ \ \ \ \ \
(c) K-map of $f_{3}$}\\
\vskip1.5mm {\small
\ \ \ \ \ Fig.\,7. K-maps of  Boolean functions\vspace{1.2mm}}
}}\end{center}

  \item The minterm canonical form of the logic expression is
 $$\left\{
\begin{array}{rcl}
x_{1}(t+1)&=&\sum m(0,2,4,6),\\
x_{2}(t+1)&=&\sum m(1,2,3,4,5,6,7),\\
x_{3}(t+1)&=&\sum m(3,7).
\end{array}\right.$$
Furthermore, the  minimization form is
 $$\left\{
\begin{array}{rcl}
x_{1}(t+1)&=&\overline{x}_{3}(t),\\
x_{2}(t+1)&=&x_{1}(t)+x_{2}(t)+x_{3}(t),\\
x_{3}(t+1)&=&{x}_{2}(t){x}_{3}(t),
\end{array}\right.$$ and the network is presented in {Figure} 2.
\end{enumerate}
\end{example}

\section{Future work}
We provide a complete solution for converting the matrix expression back to the logic expression of a BN in this paper. It is known that the complexity of computation is a series
problem in the method for converting the logic expression into the matrix
expression provided in \cite{LinearRepresentation}.
Does there exist a method with a lower computational complexity that can be used to solve this  conversion problem? And what is the lowest computational complexity for solving such a problem?  It is worth mentioning that the K-map of a BN might be useful in computing the transition matrix of a BN from its logic expression. By using the K-map of a BN,
a method with the lowest computational complexity is hopeful to be developed to convert the logic expression into the matrix
expression of a BN in a future paper. In addition, the conversion methods might be able to be extended to Boolean control networks.
The applications of our method in studying some control problems, such as the stability, controllability and observability, of Boolean control networks deserve further research.

\end{CJK}

\begin{thebibliography}{10}

\bibitem{RetrievingBoolean}
E.~Agliari, A.~Barra, C.~Longo, and D.~Tantari.
\newblock Neural networks retrieving {B}oolean patterns in a sea of {G}aussian
  ones.
\newblock {\em Journal of Statistical Physics}, 168(5):1085--1104, 2017.

\bibitem{DEDS}
W.~Chen and Y.~Tao.
\newblock Observabilities and reachabilities of nonlinear {DEDS} and coloring
  graphs.
\newblock {\em Chinese Science Bulletin}, 46(8):642--644, 2001.

\bibitem{Controllabilityobservability}
D.~Cheng and H.~Qi.
\newblock Controllability and observability of {B}oolean control networks.
\newblock {\em Automatica}, 45:1659--1667, 2009.

\bibitem{LinearRepresentation}
D.~Cheng and H.~Qi.
\newblock A linear representation of dynamics of {B}oolean networks.
\newblock {\em IEEE Transactions on Automatic Control}, 55(10):2251--2258,
  2010.

\bibitem{fixedpoint}
J.~Cochet-Terrasson, S.~Gaubert, and J.~Gunawardena.
\newblock A constructive fixed point theorem for min-max functions.
\newblock {\em Dynamics and Stability of Systems}, 14(4):407--433, 1999.

\bibitem{Attractor}
A.~A. Frolov, D.~Husek, I.~P. Muraviev, and P.~Y. Polyakov.
\newblock Boolean factor analysis by attractor neural network.
\newblock {\em IEEE Transactions on Neural Networks}, 18(3):698--707, 2007.

\bibitem{dualitytheorem}
S.~Gaubert and J.~Gunawarden.
\newblock The duality theorem for min-max functions.
\newblock {\em Comptes Rendus de l'Academie des Sciences. Series I:
  Mathematics}, 326(1):43--48, 1998.

\bibitem{Minmaxfunctions}
J.~Gunawardena.
\newblock Min-max functions.
\newblock {\em Discrete Event Dynamic Systems}, 4(4):377--407, 1994.

\bibitem{NetworkBiology}
E.~Haliki and N.~Kazanci.
\newblock Effects of a silenced gene in {B}oolean network models.
\newblock {\em Network Biology}, 7(1):10--20, 2017.

\bibitem{Kauffman}
S.~A. Kauffman.
\newblock Metabolic stability and epigenesis in randomly constructed genetic
  nets.
\newblock {\em Journal of Theoretical Biology}, 22:437--467, 1969.

\bibitem{LAC}
L.~Menini, C.~Possieri, and A.~Tornambe.
\newblock Boolean network representation of a continuous-time system and
  finite-horizon optimal control: application to the single-gene regulatory
  system for the {LAC} operon.
\newblock {\em International Journal of Control}, 90(3):535--568, 2017.

\bibitem{dynamicmaxmin}
G.~J. Olsder.
\newblock Eigenvalues of dynamic max-min systems.
\newblock {\em Discrete Event Dynamic Systems}, 1(2):177--207, 1991.

\bibitem{SocialNetwork}
G.~A. Polacek, D.~Verma, and S.~Y. Yang.
\newblock Influencing message propagation in a social network using embedded
  {B}oolean networks: A demonstration using agent-based modeling.
\newblock {\em INCOSE International Symposium}, 26(1):1509--1523, 2016.

\bibitem{LogicDesign}
C.~H. Roth and L.~L. Kinney.
\newblock {\em Fundamentals of Logic Design}.
\newblock CENGAGE Learning, 2010.

\bibitem{BooleanFunctions}
W.~G. Schneeweiss.
\newblock {\em Boolean Functions with Engineering Applications and Computer
  Programs}.
\newblock Springer-Verlag, London, 1989.

\bibitem{StateFeedback}
Y.~Tao and G.-P. Liu.
\newblock State feedback stabilization and majorizing achievement of
  min-max-plus systems.
\newblock {\em IEEE Transactions on Automatic Control}, 50(12):2027--2033,
  2005.

\bibitem{112}
Y.~Tao, G.-P. Liu, and X.~Mu.
\newblock Max-plus matrix method and cycle time assignability and feedback
  stabilizability for min-max-plus systems.
\newblock {\em Mathematics of Control, Signals, and Systems}, 25(2):197--229,
  2013.

\end{thebibliography}
\end{document}